\documentclass[twocolumn,aps,prl,amsmath,amssymb,superscriptaddress]{revtex4-2}

\usepackage{newtxtext}
\usepackage{newtxmath}
\usepackage{microtype}
\usepackage{amsmath}
\usepackage{amssymb}
\usepackage{bm}
\usepackage{braket}

\usepackage{dcolumn}
\usepackage{multirow}
\usepackage{booktabs}
\usepackage{tabularx}
\usepackage{cases}

\usepackage{graphicx}

\usepackage{enumerate}

\usepackage{xcolor}
\definecolor{PRLBlue}{rgb}{0.19,0.54,0.92}

\usepackage[
    colorlinks,
    citecolor=PRLBlue,
    linkcolor=PRLBlue,
    urlcolor=PRLBlue
]{hyperref}

\usepackage{orcidlink}

\definecolor{TighnariBrown}{RGB}{141,87,41}
\definecolor{TighnariGreen}{RGB}{40,114,70}
\definecolor{TighnariYellow}{RGB}{247,191,99}

\def\beq{\begin{equation}}
\def\eeq{\end{equation}}
\def\bald{\begin{aligned}}
\def\eald{\end{aligned}}
\def\bea{\begin{eqnarray}}
\def\eea{\end{eqnarray}}

\def\ket#1{\left|#1\right\rangle}

\def\Tr{\mathrm{Tr}}

\def\Eq#1{Eq.~(\ref{#1})}

\makeatletter
\let\oldaddcontentsline\addcontentsline
\renewcommand{\addcontentsline}[3]{}
\makeatother

\begin{document}
	\title{Universal Driven Critical Dynamics of Entanglement Entropy}
	\author{Chang-Yu Shen}
    \affiliation{Beijing National Laboratory for Condensed Matter Physics\\ \& Institute of Physics, Chinese Academy of Sciences, Beijing 100190, China}
	\affiliation{University of Chinese Academy of Sciences, Beijing 100049, China}

	\author{Shuai Yin}
    \email{yinsh6@mail.sysu.edu.cn}
    \affiliation{Guangdong Provincial Key Laboratory of Magnetoelectric Physics and Devices,\\ Sun Yat-Sen University, Guangzhou 510275, China}
    \affiliation{School of physics, Sun Yat-Sen University, Guangzhou 510275, China}
    
	\author{Zi-Xiang Li}
	\email{zixiangli@iphy.ac.cn}
    \affiliation{Beijing National Laboratory for Condensed Matter Physics\\ \& Institute of Physics, Chinese Academy of Sciences, Beijing 100190, China}
	\affiliation{University of Chinese Academy of Sciences, Beijing 100049, China}

	\date{\today}
	\begin{abstract}
		The Kibble-Zurek mechanism (KZM) and finite-time scaling (FTS) provide a foundational framework for driven critical dynamics, yet their predictive power has been largely confined to local observables. Here, we establish a universal finite-time scaling theory for the nonequilibrium dynamics of quantum entanglement. Using unbiased quantum Monte Carlo simulations, we investigate the corner entanglement entropy of (2+1)-dimensional interacting Dirac fermions driven from ordered phases toward a quantum critical point. We find that the corner entanglement accurately obeys a universal driven scaling governed by the driving rate and system size, persisting whether the initial ordered state is fully gapped or hosts gapless Goldstone modes. Crucially, this dynamical entanglement exhibits a logarithmic dependence on the driving rate, from which the universal corner coefficient of the underlying conformal field theory can be robustly extracted far from equilibrium. These results generalize the KZM from local observables to the intrinsic nonlocal quantum information measures, offering a practical blueprint for characterizing quantum criticality and entanglement on programmable quantum simulators.
	\end{abstract}
	\maketitle

    \pdfbookmark[1]{Introduction}{sec1}
    \textcolor{black}{\it Introduction}--- Unveiling the universal nonequilibrium dynamics of driven quantum many-body systems is a fundamental challenge at the intersection of condensed matter physics and quantum information science. For decades, the Kibble-Zurek mechanism (KZM) and its finite-time scaling (FTS) generalization~\cite{Kibble1976JPA,Zurek1985Nature,Pol2005prb,Zurek2005prl,Dziarmaga2005prl,Zhifangxu2005prb,Zeng2025NC,Gong_2010,Huang2014PRB,Chandran2012prb,del_Campo_2014,Polkovnikov2011RMP,Dziarmaga2010Adv} have served as the theoretical cornerstone in this field, dictating how universal defect formation emerges when a system is ramped across a quantum critical point (QCP). Today, this framework is undergoing a rapid renaissance fueled by programmable quantum simulators. Platforms such as Rydberg atom arrays~\cite{keesling2019nature,ebadi2021nature,Chen2025PRLKZ}, ultracold atoms in optical lattices~\cite{bloch2012np}, and trapped ions~\cite{Ulm2013NC,Pyka2013NC} have not only achieved high-precision validations of KZM physics, but have also demonstrated the unprecedented capability to directly measure quantum entanglement in out-of-equilibrium settings~\cite{Elben2023,Brydges2019,Greiner2015NatureEE}. These experimental milestones pose a compelling theoretical mandate: extending the powerful paradigm of universal driven scaling beyond local observables and into the intrinsic quantum entanglement of the many-body wavefunction.

	To date, however, the predictive power of the KZM and FTS paradigms has been almost exclusively tested against \emph{local} observables, such as defect densities and order-parameter correlations. Extending this framework requires a genuinely non-local quantum information measure, for which entanglement entropy serves as the quintessential diagnostic tool~\cite{RMPEntanglement}. While the driven dynamics of entanglement entropy has been extensively explored in one-dimensional systems~\cite{Zurek2007PRA,Yin2016PRBdriven,cpl_42_11_110001}, where universal criticality is robustly captured by the leading logarithmic term of (1+1)D conformal field theory (CFT), extending this paradigm to (2+1)D is highly non-trivial, since the entanglement structure in two dimensions bears fundamental differences from its one‑dimensional counterpart. In (2+1)D, entanglement entropy is dominated by the non-universal area law, and the intrinsic quantum critical data is instead delicately encoded in a subleading logarithmic correction arising from sharp corners in the subsystem boundary~\cite{Fradkin2006PRL,Casini2007NPB,TomoyoshiHirata_2007,Bueno2019}:
\begin{equation}
    S_2 = bL - a_c(\theta)\ln L + \text{const.}
    \label{eq_intro}
\end{equation}
Crucially, the angle-dependent corner coefficient $a_c(\theta)$ is dictated solely by the universality class of the underlying CFT. It encodes intrinsic universal data that remains inaccessible to conventional local probes. While these equilibrium properties are well established, the behavior of entanglement entropy under driven dynamics in (2+1)D remains a completely uncharted frontier, posing a fundamental open question: does the universal information of quantum entanglement survive the nonequilibrium driving process, and if so, what scaling laws govern its behavior?

	\begin{figure}[!t]
		\centering
		\includegraphics[width=\linewidth]{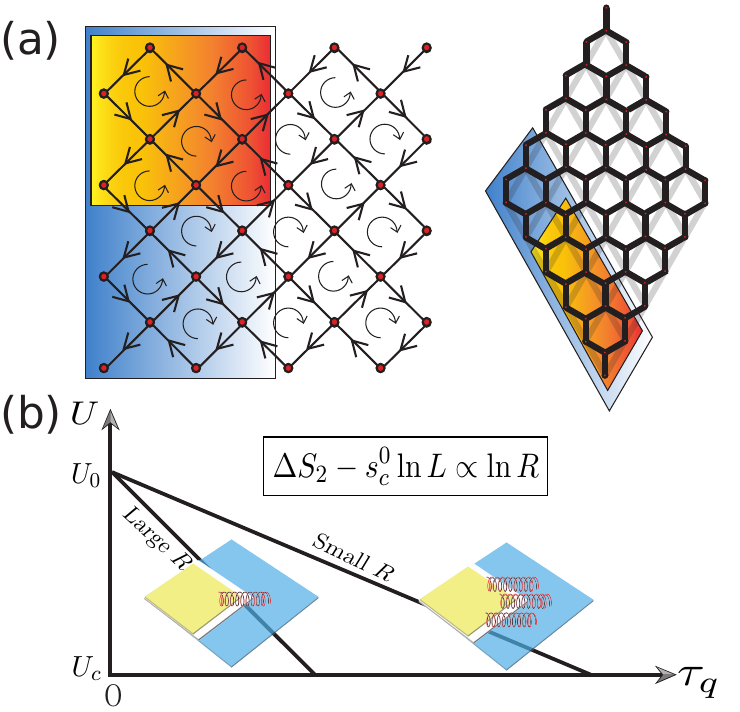}
		\caption{\textbf{(a)} 
            Lattices and entanglement-entropy subsystem cuts considered in this Letter. The full system has periodic boundary conditions. On the square lattice, blue Region \(A\) (\(\frac{L}{2}\times L\)) has a smooth boundary, while orange Region \(B\) (\(\frac{L}{2}\times\frac{L}{2}\)) has four \(\frac{\pi}{2}\) corners. On the honeycomb lattice, the blue Region \(A\) (\(\frac{L}{3}\times L\)) has a smooth boundary, whereas the orange Region \(B\) (\(\frac{L}{3}\times\frac{2L}{3}\)) has two \(\frac{\pi}{3}\) and two \(\frac{2\pi}{3}\) corners. In each case, the two regions have equal boundary lengths.
			\textbf{(b)} Schematic illustration of the linear driving protocol. The initial state is the ground state of the Hamiltonian with interaction strength $U_0$ deep in the ordered phase. In the driven process, the interaction is linearly ramped from $U_0$ toward the quantum critical point $U_c$ over imaginary time $\tau_q$, with driving rate $R\equiv|U_c-U_0|/\tau_q$. The corner entanglement entropy $\Delta S_2$ is measured at the endpoint of the drive.
		}
		\label{SCEE_png}
	\end{figure}

    In this work, we address these questions by studying the corner contribution to the entanglement entropy under external driving in Dirac fermion systems. Using unbiased quantum Monte Carlo (QMC) simulations, we investigate interacting lattice models driven from the ordered phase toward the QCP, covering both gapped and gapless initial states within a unified framework. We reveal that the corner entanglement entropy $\Delta S_2$ accurately obeys a universal FTS form, governed solely by the competition between the driving rate $R$ and the system size $L$. Crucially, $\Delta S_2$ exhibits a logarithmic dependence on the driving rate, with its prefactor directly encoding the universal corner coefficient of the underlying CFT. This scaling structure holds irrespective of whether the initial ordered phase is gapless (hosting Goldstone modes) or fully gapped. By leveraging this driven scaling, we successfully extract the universal corner coefficient $s_c^{\mathrm{QCP}}$ characterizing the QCP. Ultimately, our findings demonstrate that the underlying conformal data of a QCP robustly survives non-adiabatic driving, thereby elevating the KZM and FTS paradigms from the level of local order parameters to the non-local entanglement structure of the QCP, and providing vital theoretical guidance for exploring the nonequilibrium properties of quantum entanglement on modern quantum simulation platforms.

    \pdfbookmark[1]{Model and Method}{sec1}
	\textcolor{black}{\it Model and Method}--- We study two different half-filled interacting Dirac-fermion models, hosting Dirac QCPs with two different universality classes, 
	defined on two lattice geometries: the honeycomb lattice and the $\pi$-flux square lattice. The first is the standard Hubbard model:
	\begin{equation}
		H = -t \sum_{\langle i,j\rangle,\sigma}
		e^{i\phi_{ij}} c_{i\sigma}^\dagger c_{j\sigma}	+ \mathrm{H.c.} + U \sum_i \left(n_{i\uparrow} - \tfrac{1}{2}\right)\left(n_{i\downarrow} - \tfrac{1}{2}\right)
	\end{equation}
	where $\phi_{ij}=0$ for the honeycomb lattice, and 
	$
	\sum_{\square} \phi_{ij} = \pi 
	$
	for the $\pi$-flux square lattice, thereby producing two Dirac fermions at low energy for each spin flavor. Although the two lattices differ microscopically, they realize the same
    class of Dirac QCP. The Hubbard interaction drives the Dirac semimetal into an antiferromagnetic (AFM) Mott insulating phase. The AFM order spontaneously breaks the SU(2) spin symmetry, leading to $N_G=2$ gapless Goldstone modes~\cite{SSB_scaling}. The corresponding QCP belongs to the chiral Heisenberg Gross-Neveu-Yukawa (GNY) universality class~\cite{Rosenstein1993PLB,Herbut2006PRL,Assaad2013prx,Sorella2012SP,Sorella2016prx,li2017nc,Scherer2017PRD,Li2018SA,Lang2019prl,Abolhassan2022PRL,Li2024PRLNon-Hermitian}.

	The second model is the spinless t-V model, including the nearest-neighbor density interaction
	\begin{equation}
    H = -t \sum_{\langle i,j\rangle}
		\left(e^{i\phi_{ij}} c_{i}^\dagger c_{j}+ \mathrm{H.c.}\right) + V \sum_{\langle i,j \rangle} \left(n_i - \tfrac{1}{2}\right)\left(n_j - \tfrac{1}{2}\right)
		\label{tV}
	\end{equation}
	where the interaction drives a transition into a charge-density-wave (CDW) ordered phase. In this case a discrete $\mathbb{Z}_2$ symmetry is broken, and therefore no Goldstone modes emerge. The corresponding QCP belongs to the chiral Ising GNY universality class. The critical interaction strengths ($U_c$ and $V_c$) defining the QCPs for these four models—encompassing both interaction types across the two lattice geometries—are summarized in Table~\ref{tab_models}.

	The simulations are performed using the projective determinant QMC method~\cite{Sorella1989EPL,AssaadReview,Assaad2025ALF,li2019arcmp}. The R\'enyi entanglement entropy of strongly correlated systems can be computed directly in QMC simulations using the replica trick, both for bosonic and spin systems in stochastic series expansion~\cite{Hastings2010,Melko2011NP,Humeniuk2012,Emidio2020PRL,Yan2025NC} and for interacting fermions in determinantal QMC~\cite{Grover2013PRL,Jiang2025arXiv,Xu2025NC}. The corner contribution to the entanglement entropy is computed with high precision using the incremental Subtracted Corner Entanglement Entropy (SCEE) algorithm~\cite{Emidio2024PRL,Liao2025npjQI,DaliaoYuan2023PRB,kups7666,Xu2024PRLDisorderoperator,DaLiaoYuan2024PRB}. Owing to the time-reversal symmetries in complex-fermion and Majorana representations, all four models are free from the fermion sign problem~\cite{Troyer2005sign,Wu2005PRBsign,Li2015PRBsign,Li2016PRLsign,xiang2016prl,wang2015prl,Yu2026SA}. Additional details are provided in the Supplementary Material (SM)~\cite{SM}.

    \pdfbookmark[1]{Finite-time scaling theory for entanglement entropy}{sec1}
	\textcolor{black}{\it Finite-time scaling theory for entanglement entropy}--- The corner contribution to the entanglement entropy is isolated by subtracting the second R\'enyi entropies of two subregions that share the same boundary length but have different geometries, defined as $\Delta S_2 = S_2^A-S_2^B$ with region $A$ (smooth) and $B$ (with corners), as illustrated in Fig.~\ref{SCEE_png}(a). In equilibrium, this subtraction removes the non-universal area-law term,
	leaving a universal corner contribution that scales logarithmically with the 
	linear system size: $\Delta S_2 =s_c \ln L + \text{const.}$, where the corner 
	coefficient $s_c$ encodes the universal information~\cite{Fradkin2006PRL,Casini2007NPB,TomoyoshiHirata_2007,Bueno2019}. This logarithmic scaling 
	form holds for both the QCP and ordered phases with gapless Goldstone modes. The total universal coefficient $s_c$ is determined by the sum of contributions from each corner angle in the subsystem. Specifically, as depicted in Fig.~\ref{SCEE_png}(a), $s_c = 2 \times [a_c(\frac{\pi}{3}) + a_c(\frac{2\pi}{3})]$ on the honeycomb lattice, and $s_c = 4 \times a_c(\frac{\pi}{2})$ on the square lattice. A detailed review of entanglement entropy scaling in (2+1)D is provided in the SM~\cite{SM}.

    Directly simulating real-time dynamics using unbiased QMC methods is severely hindered by the dynamical sign problem. Remarkably, scaling analyses demonstrate that observables under both real- and imaginary-time driven dynamics share identical universal scaling forms~\cite{Grandi2011,YinShuai2014PRB}. This rigorous correspondence, verified in various previous studies~\cite{Zeng2025NC,Yu2026PRL,Yin2022prl,Yin2025nc,Shu2022prb,Shu2017prb}, provides a powerful, sign-problem-free framework to extract the critical scaling behavior of driven dynamics via imaginary-time evolution. We employ a linear driving protocol, as illustrated in Fig.~\ref{SCEE_png}(b): the system is prepared in the ground state at $U_0$ deep within the ordered phase, and then driven toward the QCP $U_c$ by linearly ramping the coupling over an imaginary-time duration $\tau_q$, defining a driving rate $R\equiv|U_c-U_0|/\tau_q$. During the ramp, the state evolves according to the imaginary-time Schrödinger equation $-\frac{\partial}{\partial\tau}|\psi(\tau)\rangle = H(\tau)|\psi(\tau)\rangle$ with the normalization condition imposed.
    
    Under the renormalization-group transformations $(U-U_c)\to(U-U_c)b^{1/\nu}$ and $\tau\to\tau b^{-z}$, the driving rate scales as $R\to Rb^r$ with $r=z+1/\nu$. This introduces an emergent nonequilibrium length scale $\xi_R\sim R^{-1/r}$ that fundamentally competes with the system size $L$~\cite{YinShuai2014PRB,Zeng2025NC,Gong_2010,Yin2016PRBdriven}. Extending the equilibrium scaling of the corner entanglement entropy to driven dynamics, we propose that at the critical point
	\begin{equation}
		\Delta S_2(R,L)=s^{\rm QCP}_c\ln L+f(RL^r)+\text{const.},
		\label{neq_scaling}
	\end{equation}
	where $f(x)$ is a universal scaling function of $x\equiv RL^r$.
	In the adiabatic limit $R\to0$, the system approaches the equilibrium
	critical state, so that $f(x)$ tends to a constant and
	$\Delta S_2=s_c^{\rm QCP}\ln L+\text{const.}$ In the large-$x$ FTS
	regime, the KZM correlation length $\xi_R$ becomes smaller
	than the system size, so the dynamics is governed by the driving
	rate rather than by finite-size effects. Meanwhile, $\xi_R$ must remain larger than the microscopic length scale $a$. Within this scaling window $a\ll\xi_R\ll L$~\cite{Zeng2025NC}, matching the logarithmic system-size
	dependence inherited from the initial state gives
	$f(x)=\frac{s^{0}_c-s^{\rm QCP}_c}{r}\ln x$, which, substituted into
	\Eq{neq_scaling}, yields our central prediction
	\begin{equation}
		\Delta S_2(R,L)=s^0_c\ln L+\frac{\Delta s_c}{r}\ln R+\text{const.}
		\label{scaling_main}
	\end{equation}
	Here $\Delta s_c\equiv s^0_c-s^{\rm QCP}_c$ denotes the change of the corner coefficient across the drive, from the initial ordered state to the QCP. The system-size dependence of $\Delta S_2$ is thus governed solely by the initial-state coefficient $s^0_c$, while the logarithmic-in-$R$ term is determined by the critical exponents through $r=z+1/\nu$.
	
	\begin{figure}[!t]
		\centering
		\includegraphics[width=\linewidth]{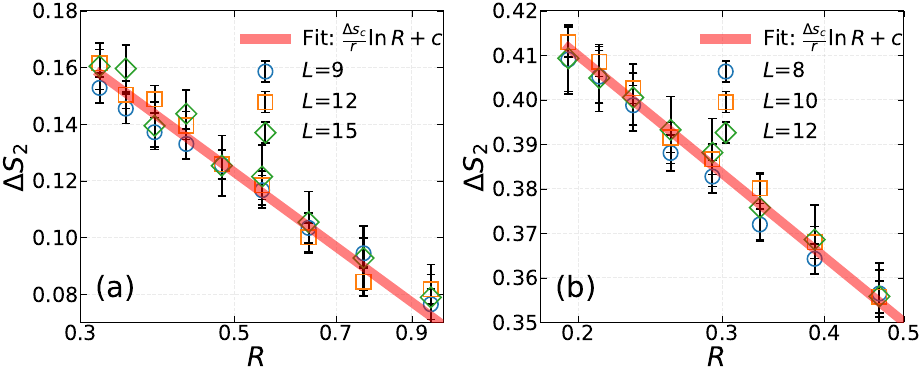}
		\caption{\textbf{Driven dynamics of the corner entanglement entropy for the chiral Ising GNY universality class.}
			\textbf{(a)} Corner entanglement entropy $\Delta S_2$ as a function of the driving rate $R$ for the $t$-$V$ model on the honeycomb lattice ($V_c\approx 1.35$). Since $s_c^0=0$ in the gapped CDW phase, $\Delta S_2$ is independent of the system size $L$: the data points at each fixed $R$ coincide for different $L$, without any subtraction, following $\Delta S_2=-(s_c^{\rm QCP}/r)\ln R+\text{const.}$ A linear fit in $\ln R$ then yields $s_c^{\rm QCP}=0.168(9)$.
			\textbf{(b)} Same analysis as in (a) for the $t$-$V$ model on the $\pi$-flux square lattice ($V_c\approx 1.30$), yielding $s_c^{\rm QCP}=0.140(7)$.
		}
		\label{fig_GNY_I}
	\end{figure}

    \pdfbookmark[1]{Driven dynamics of chiral Ising universality class}{sec1}
	\textcolor{black}{\it Driven dynamics of the chiral Ising universality class.}---We begin our analysis with the two spinless $t$-$V$ models. The systems are initially prepared deep in the CDW ordered phase---at $V_0=2.5$ and $V_0=2.0$ for the honeycomb and $\pi$-flux square lattices, respectively---and then driven toward the chiral Ising GNY quantum critical point~\cite{Li2015njp,Wang2014njp}. Because the CDW phase spontaneously breaks only a discrete $\mathbb{Z}_2$ symmetry, it is fully gapped and devoid of Goldstone modes. Consequently, the initial corner entanglement coefficient vanishes exactly, $s^0_c=0$. This allows the nonequilibrium logarithmic dependence of $\Delta S_2$ on the driving rate $R$ to be isolated without requiring any background subtraction.

	Fig.~\ref{fig_GNY_I} elegantly demonstrates this behavior. The endpoint corner entanglement entropy $\Delta S_2$ is plotted as a function of $R$ for both the honeycomb [Fig.~\ref{fig_GNY_I}(a)] and $\pi$-flux square [Fig.~\ref{fig_GNY_I}(b)] lattices. With $s^0_c=0$, our central analytical prediction in \Eq{scaling_main} simplifies dramatically to:
\begin{equation}
	\Delta S_2(R,L)=-\frac{s^{\rm QCP}_c}{r}\ln R+\text{const.},
	\label{scaling_ising}
\end{equation}
This simplified scaling relation dictates two stark consequences. First, $\Delta S_2$ becomes strictly independent of the system size $L$: data points at a fixed $R$ perfectly collapse onto a single curve across all simulated $L$. Second, $\Delta S_2$ exhibits a clean linear dependence on $\ln R$, from whose slope the universal corner coefficient $s_c^{\rm QCP}$ at the QCP can be directly extracted. This pristine data collapse represents the most transparent manifestation of the driven scaling theory. Physically, this remarkable behavior can be intuitively understood through the lens of the KZM. In the nonequilibrium scaling regime ($RL^r \gg 1$), the relevant characteristic length scale of the system is no longer the spatial size $L$, but rather the emergent KZM correlation length $\xi_R \sim R^{-1/r}$. Substituting $L \to \xi_R$ directly into the equilibrium critical scaling relation $\Delta S_2 \sim s_c^{\mathrm{QCP}} \ln L$ elegantly yields the observed $-\frac{s_c^{\mathrm{QCP}}}{r} \ln R$ dependence. This provides a compelling physical picture that explicitly bridges the spatial conformal data of the QCP with the temporal driving rate.

	By combining the slope of the logarithmic scaling in \Eq{scaling_ising} with the known correlation-length exponent $\nu=0.87(6)$~\cite{wang2026} for the chiral Ising universality class, we extract $s_c^{\rm QCP}=0.168(9)$ for the honeycomb lattice and $s_c^{\rm QCP}=0.140(7)$ for the $\pi$-flux square lattice. This represents the first determination of the universal corner entanglement coefficient for the chiral Ising GNY QCP. Crucially, the extracted values of $s_c^{\rm QCP}$ on both lattice geometries are significantly larger than those of free Dirac fermions (see Table~\ref{tab_models}). This pronounced deviation reveals that the critical entanglement is fundamentally enhanced by the strong coupling between the Dirac fermions and the fluctuating order-parameter bosons. 
    
    \begin{figure*}[!t]
    	\centering
    	\includegraphics[width=\linewidth]{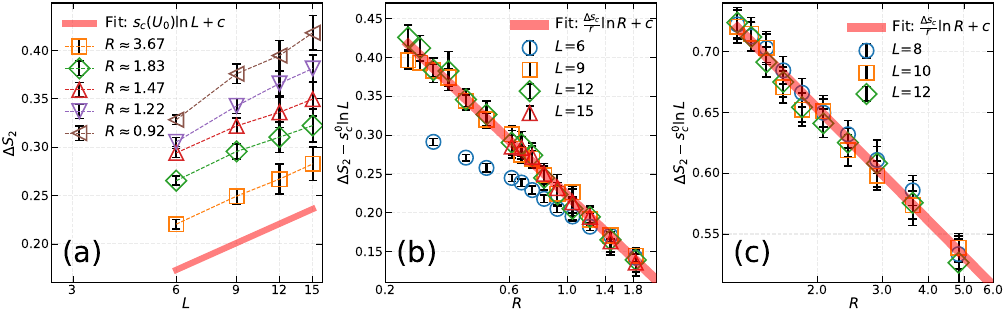}
    	\caption{\textbf{Driven dynamics of the corner entanglement entropy for the chiral Heisenberg GNY universality class.}
    		\textbf{(a)} System-size dependence of the corner entanglement entropy $\Delta S_2$ at several fixed driving rates $R$ for the Hubbard model on the honeycomb lattice. The data exhibit a logarithmic scaling behavior
    		$
    		\Delta S_2 \sim s_c^0\ln L ,
    		$ indicating that, at finite driving rates, the entanglement entropy remains dominated by the corner contribution inherited from the initial AFM ordered phase. Solid lines represent logarithmic fits.
            \textbf{(b)} Shifted corner entanglement entropy $\Delta S_2-s_c^0\ln L$ as a function of the driving rate $R$ for the Hubbard model on the honeycomb lattice, with the initial-state corner coefficient $s_c^0=0.0688$. After subtracting the initial-state contribution, data for different system sizes collapse onto a single curve, demonstrating the universal driven scaling behavior. The collapsed data show a linear dependence on $\ln R$, from whose slope the corner coefficient at the quantum critical point is extracted as $s_c^{\rm QCP}=0.332(11)$.
            \textbf{(c)} Same analysis as in (b) for the Hubbard model on the $\pi$-flux square lattice, with $s_c^0=0.052$. The data collapse onto an analogous universal scaling curve, yielding $s_c^{\rm QCP}=0.304(12)$. The consistency of the driven scaling behavior across the two lattice geometries confirms its universality within the chiral Heisenberg GNY class.}
    	\label{fig_GNY_H}
    \end{figure*}

    \pdfbookmark[1]{Driven dynamics of chiral Heisenberg universality class}{sec1}
	\textcolor{black}{\it Driven dynamics of chiral Heisenberg universality class}--- We now turn to the Hubbard models. The initial states are prepared deep in the AFM ordered phase---at $U_0=6.0$ for the honeycomb lattice and $U_0=10.0$ for the $\pi$-flux square lattice---before being driven toward the chiral Heisenberg GNY QCP. Unlike the fully gapped CDW state, the AFM phase spontaneously breaks the continuous SU(2) spin-rotational symmetry, thereby hosting $N_G=2$ gapless Goldstone modes. Consequently, the initial corner entanglement coefficient is strictly non-zero, yielding $s_c^0=0.0688$ and $0.0520$ for the honeycomb and $\pi$-flux square lattices, respectively (see Table~\ref{tab_models})~\cite{Helmes2016PRB}. Crucially, a conventional prerequisite for the KZM and FTS paradigms is the existence of a finite energy gap in the initial state. While a generalized FTS framework for gapless initial states was recently established for local observables~\cite{Zeng2025NC}, its applicability to non-local quantum information measures remains completely unknown. Determining whether the universal driven scaling of entanglement entropy survives the presence of gapless initial modes thus constitutes a fundamental open question.

	Fig.~\ref{fig_GNY_H}(a) shows the system-size dependence of the subtracted corner entanglement entropy $\Delta S_2$ on the honeycomb lattice at several fixed driving rates $R$. For $L\ge 9$, the data exhibit a clean logarithmic scaling $\Delta S_2\sim s_c^0\ln L$, confirming that at finite driving rates the entanglement entropy remains dominated by the corner contribution inherited from the initial AFM phase. The $L=6$ data deviate slightly from this logarithmic trend due to finite-size effects, as expected for the smallest system size. Fig.~\ref{fig_GNY_H}(b) and Fig.~\ref{fig_GNY_H}(c) present the central scaling analysis for the honeycomb and $\pi$-flux square lattices, respectively. By subtracting the initial-state Goldstone contribution---evaluating $\Delta S_2-s_c^0\ln L$---we isolate the purely dynamical nonequilibrium part. When plotted against the driving rate $R$, the data for $L\ge 9$ collapse onto a single universal master curve for each lattice, while the $L=6$ data show visible deviations, again attributable to finite-size corrections. This master curve smoothly captures the linear regime in $\ln R$, exactly as predicted by \Eq{scaling_main}.

	From the slope of this linear regime and the independently known critical exponent $\nu=1.11(9)$~\cite{wang2026,Li2015njp,Sorella2016prx}, we extract the universal corner coefficient at the chiral-Heisenberg GNY QCP: $s_c^{\rm QCP}=0.332(11)$ for the honeycomb lattice and $s_c^{\rm QCP}=0.304(12)$ for the $\pi$-flux square lattice. These values are consistent with previous equilibrium and short-time relaxation results~\cite{Emidio2024PRL,shen2026}, confirming that the driven scaling theory correctly captures the universal entanglement properties of the QCP despite the gapless nature of the initial state.

    Across all four models---encompassing two symmetry classes, two lattice geometries, and two fundamentally distinct initial states---the corner entanglement entropy is rigorously governed by the universal scaling relation of \Eq{scaling_main}. Furthermore, the quantitative agreement between our dynamically extracted QCP corner coefficients and previous equilibrium results firmly establishes the driven scaling approach as a reliable probe for universal quantum-information data. Crucially, this unified scaling behavior of quantum entanglement persists regardless of the gapped or gapless nature of the initial ordered state, highlighting the extraordinary universality of nonequilibrium entanglement dynamics.

		\begin{table*}[ht!]
		\centering
        \begin{ruledtabular}
		\begin{tabular}{@{}cccccccccc@{}}
			\addlinespace[0.em]
			Model & GNY class & Lattice & $U_c/V_c$ & $N_\mathrm{f}$ & $N_G$ & $s_c^{0}$ & $\nu$ & $s_c^{\text{f}}$ & $s_c^{\text{QCP}}$ \\
			\addlinespace[0.0em]
			\hline
			\addlinespace[0.1em]
			\multirow{2}{*}{Hubbard} & \multirow{2}{*}{chiral Heisenberg} & honeycomb & $U_c\approx 3.80$ & 4 & 2 & 0.0688 & 1.11(9) & 0.3112 &  0.332(11) \\
			& & $\pi$-flux square & $U_c\approx 5.65$ & 4 & 2 & 0.0520 & 1.11(9) & 0.2396 &  0.304(12) \\
			\addlinespace[0.05em]
			\hline
			\addlinespace[0.1em]
			\multirow{2}{*}{t-V} & \multirow{2}{*}{chiral Ising} & honeycomb & $V_c\approx 1.35$ & 2 & 0 & 0 & 0.87(6) & 0.1556 & 0.168(9) \\
			& & $\pi$-flux square & $V_c\approx 1.30$ & 2 & 0 & 0 & 0.87(6) & 0.1198 & 0.140(7) \\
			\addlinespace[-0.2em]
		\end{tabular}
		\caption{Summary of the four interacting Dirac-fermion models in this work: the Hubbard and spinless $t$-$V$ models on both the honeycomb and $\pi$-flux square lattices. $N_\mathrm{f}$ denotes the number of free Dirac fermions in the low-energy effective theory, and $N_G$ is the number of Goldstone modes in the initial ordered phase. The ordered-state corner entanglement coefficient $s_c^{0}$~\cite{Helmes2016PRB} vanishes exactly for the fully gapped CDW phase where $N_G=0$. The baseline free-fermion corner contribution is given by $s_c^{\text{f}}=N_\mathrm{f}\times s_c^{(\text{1 fermion})}$, where $s_c^{(\text{1 fermion})}$ is the analytically known~\cite{Casini2007NPB,Helmes2016PRB}. Finally, $s_c^{\rm QCP}$ is the universal corner coefficient at the QCP, dynamically extracted from our driven scaling analysis using the established correlation-length exponent $\nu$ from independent equilibrium studies of local order parameters~\cite{Li2015njp,Sorella2016prx,wang2026}.}
		\label{tab_models}
        \end{ruledtabular}
	\end{table*}

    \pdfbookmark[1]{Discussions and concluding remarks}{sec1}
	\textcolor{black}{\it Discussions and concluding remarks}--- In summary, we have established a definitive conceptual link between the finite-time scaling paradigm and the nonequilibrium dynamics of quantum entanglement. Through numerically-exact simulations of four distinct (2+1)D Dirac fermion models across two fundamental universality classes---the chiral Heisenberg and chiral Ising GNY QCPs---we have demonstrated that the corner entanglement entropy $\Delta S_2$ flawlessly obeys a universal driven scaling. This confirms that the universal KZM and the associated FTS theory extend far beyond local observables, dictating the dynamical evolution of non-local entanglement irrespective of any microscopic or geometric details.

	Our findings culminate in two principal implications. First, we establish the first universal driven scaling theory for quantum entanglement in (2+1)D critical systems. This profoundly enriches the fundamental framework of nonequilibrium dynamics, revealing that the intricate universal footprint of quantum entanglement is encoded in the behavior of non-adiabatic driving.  We demonstrate that this unified driven scaling structure dictates the entanglement evolution regardless of whether the initial phase is fully gapped or gapless. This structural robustness extends beyond the traditional prerequisites of the KZM, confirming that the generalized FTS, recently established for local observables~\cite{Zeng2025NC}, is a much deeper phenomenon governing the intrinsic entanglement of the many-body wavefunction. Second, and perhaps most crucially, our theory carries vital practical significance for the experimental exploration of quantum criticality. Because strict adiabatic state preparation is severely hindered by the critical slowing down inherent to quantum simulators, dynamic driving protocols have become the indispensable paradigm for investigating critical phenomena. Our work provides the theoretical foundation for extending this approach into the realm of quantum information. By demonstrating that universal CFT data can be extracted directly from the finite-rate driving of entanglement, we establish a rigorous blueprint to measure and interpret quantum critical entanglement on modern programmable simulators. Experimentally, our proposal can be readily implemented using randomized measurement protocols~\cite{Elben2023,Brydges2019} or twin-state interferometry~\cite{Greiner2015NatureEE}, both of which have already successfully probed Rényi entanglement entropies.

    Looking forward, the universality of our theoretical framework invites broad explorations across diverse physical platforms. While our current demonstrations focus on Dirac fermions, the underlying scaling principles are fundamentally rooted in conformal field theory, guaranteeing their direct applicability to interacting spin and bosonic systems. Furthermore, this dynamic scaling approach opens a compelling new pathway for unveiling exotic quantum phase transitions that transcend the conventional paradigm. Most notably, applying this framework to deconfined quantum critical points~\cite{Senthil_2004,Sandvik2007}---where local probes often struggle to unambiguously capture emergent fractionalization and gauge fields---could provide the definitive non-local signatures needed to resolve long-standing theoretical controversies. 

    \pdfbookmark[1]{Acknowledgments}{sec1}
	\textcolor{black}{\it Acknowledgments}--- C.Y.S. and Z.X.L. are supported by the National Natural Science Foundation of China under Grant Nos. 12347107 and 12474146, and Beijing Natural Science Foundation under Grant No. JR25007. S.Y. is supported by the National Natural Science Foundation of China (Grant No. 12222515), the Research Center for Magnetoelectric Physics of Guangdong Province (Grant No. 2024B0303390001), and the Guangdong Provincial Key Laboratory of Magnetoelectric Physics and Devices (Grant No. 2022B1212010008).

    \pdfbookmark[1]{Data availability}{sec1}
	\textcolor{black}{\it Data availability}--- 
	The data that support the findings of this study are available from the corresponding authors (Z.X.L. and S.Y.) upon request.

\onecolumngrid
\newpage
\widetext
\thispagestyle{empty}

\makeatletter
\let\addcontentsline\oldaddcontentsline
\makeatother

\setcounter{equation}{0}
\setcounter{figure}{0}
\setcounter{table}{0}

\renewcommand\floatpagefraction{0.9}
\renewcommand\textfraction{0.1}
\renewcommand{\theequation}{S\arabic{equation}}
\renewcommand{\thefigure}{S\arabic{figure}}
\renewcommand{\thetable}{S\arabic{table}}
\renewcommand{\theHequation}{S\arabic{equation}}
\renewcommand{\theHfigure}{S\arabic{figure}}
\renewcommand{\theHtable}{S\arabic{table}}

\pdfbookmark[0]{Supplementary Materials}{SM}
\begin{center}
    \vspace{3em}
    {\Large\textbf{Supplementary Materials for}}\\
    \vspace{1em}
    {\large\textbf{Universal Driven Critical Dynamics of Entanglement Entropy}}\\
    \vspace{0.5em}
\end{center}

\tableofcontents

\section{I. Entanglement entropy scaling in (2+1)D quantum critical systems}

In (2+1)-dimensional quantum critical systems described by a conformal field theory (CFT), the second R\'enyi entanglement entropy of a subregion with linear size $L$ follows the general scaling form~\cite{Fradkin2006PRL,Casini2007NPB,TomoyoshiHirata_2007,Bueno2019}:
\begin{equation}
	S_2^A(L) = aL - s_c(\theta)\ln L + \text{const.} + \mathcal{O}(1/L).
	\label{SM_scalingeq}
\end{equation}
The leading term $aL$ is the area-law (perimeter-law in 2D) contribution, where the coefficient $a$ is non-universal and depends on microscopic details such as the lattice structure and UV cutoff. The universal information is encoded in the subleading logarithmic correction, which arises from sharp corners in the entangling boundary. The angle-dependent coefficient $s_c(\theta)$ is determined solely by the universality class of the quantum critical system and the opening angle $\theta$ of the corner. For a smooth boundary (no corners), this logarithmic term is absent.

When the boundary contains multiple corners, the total corner coefficient is the sum of individual contributions:
\begin{equation}
	s_c = \sum_i s_c(\theta_i) = N_{\mathrm{f/b}} \sum_i  a_2^{\mathrm{f/b}}(\theta_i),
	\label{SM_sc_sum}
\end{equation}
where $a_2^{\mathrm{f/b}}(\theta)$ are the universal angle-dependent coefficients for free Dirac fermions or real bosons determined by the underlying CFT, and $N_{\mathrm{f/b}}$ denotes the number of fermion or boson species. The exact corner functions $a_2^{\mathrm{f}}(\theta)$ for free Dirac fermions and $a_2^{\mathrm{b}}(\theta)$ for free real bosons have been computed analytically and verified numerically~\cite{Casini2007NPB,Helmes2016PRB}.

In phases with spontaneous breaking of a continuous symmetry, an additional logarithmic contribution from Goldstone modes appears~\cite{SSB_scaling}:
\begin{equation}
	S_2^A(L) = aL - (s_G + s_c(\theta))\ln L + \text{const.} + \mathcal{O}(1/L),
	\label{SM_SSB_scaling}
\end{equation}
where $s_G = -N_G/2$ and $N_G$ is the number of Goldstone modes. This Goldstone contribution $s_G$ depends only on the bulk properties of the symmetry-broken phase and is independent of the boundary geometry. In the SU(2) Hubbard models studied in this work, the AFM ordered phase has $N_G=2$ Goldstone modes, while in the $t$-$V$ models the CDW phase breaks a discrete $\mathbb{Z}_2$ symmetry and has no Goldstone modes ($N_G=0$).

In contrast, for gapped ordered phases such as the CDW state with discrete $\mathbb{Z}_2$ symmetry breaking, the bulk excitation spectrum is fully gapped and all correlations decay exponentially. As a result, the entanglement entropy contains no universal logarithmic correction:
\begin{equation}
	S_2^A(L) = aL + \text{const.} + \mathcal{O}(e^{-L/\xi}),
	\label{SM_gapped_scaling}
\end{equation}
where $\xi$ is the finite correlation length. In this case $s_c=0$ and $s_G=0$, so the logarithmic coefficient vanishes and the SCEE approaches an $L$-independent constant, $\Delta S_2\to\text{const.}$, in the thermodynamic limit. This constant need not be zero because the two subsystem geometries can have different nonuniversal constant terms. This distinction between the gapless AFM initial state ($s_c\neq 0$ due to Goldstone bosons) and the gapped CDW initial state ($s_c=0$) is central to the comparison made in this work.

\textit{Subtracted corner entanglement entropy (SCEE).}---To isolate the universal corner contribution from the dominant non-universal area-law term, we employ the SCEE method~\cite{kups7666,Xu2024PRLDisorderoperator,DaLiaoYuan2024PRB}. Two entanglement regions with identical boundary lengths but different geometries are defined: region $A$ has a smooth (corner-free) boundary, while region $B$ contains sharp corners. The SCEE is defined as
\begin{equation}
	\Delta S_2 \equiv S_2^A - S_2^B.
	\label{SM_SCEE_def}
\end{equation}
Since both regions share the same boundary length, the leading area-law terms cancel exactly. In symmetry-broken phases, the Goldstone-mode contribution $s_G\ln L$, which depends only on system size and not on geometry, also cancels in the subtraction. The remaining leading contribution is purely the universal corner term:
\begin{equation}
	\Delta S_2 = s_c\ln L + \text{const.} + \mathcal{O}(1/L).
	\label{SM_SCEE_scaling}
\end{equation}

In our simulations, the subsystem geometries are chosen as follows. On the $\pi$-flux square lattice, region $A$ has dimensions $\frac{L}{2}\times L$ with a smooth boundary, and region $B$ has dimensions $\frac{L}{2}\times\frac{L}{2}$ with four $\frac{\pi}{2}$ corners. On the honeycomb lattice, region $A$ has dimensions $\frac{L}{3}\times L$ with a smooth boundary, and region $B$ has dimensions $\frac{L}{3}\times\frac{2L}{3}$ with two $\frac{\pi}{3}$ and two $\frac{2\pi}{3}$ corners.

Table~\ref{tab_sc_species} summarizes the angle-dependent corner functions $a_2^{\mathrm{f}}(\theta)$ and $a_2^{\mathrm{b}}(\theta)$ for a single two-component complex Dirac fermion and a single real boson, respectively, at the three corner angles relevant to our subsystem geometries~\cite{Helmes2016PRB}. The low-energy field content of the corresponding GNY critical points can be represented schematically by the Euclidean Lagrangian~\cite{Rosenstein1993PLB,Herbut2006PRL}
\begin{equation}
	\mathcal{L}_{\mathrm{GNY}}
	= \bar{\Psi}\gamma_\mu \partial_\mu \Psi
	+ \frac{1}{2}\sum_{a=1}^{N_\mathrm{b}}\left[(\partial_\mu\phi_a)^2+m_\phi^2\phi_a^2\right]
	+ \frac{\lambda}{4!}\left(\sum_{a=1}^{N_\mathrm{b}}\phi_a^2\right)^2
	+ g\sum_{a=1}^{N_\mathrm{b}}\phi_a\bar{\Psi}M_a\Psi,
	\label{SM_GNY_Lagrangian}
\end{equation}
where $\Psi=(\psi_1,\ldots,\psi_{N_\mathrm{f}})^T$ is the Dirac multiplet, each $\psi_i$ is a two-component complex Dirac field, $\phi_a$ is a real order-parameter field, and $M_a$ specifies the corresponding fermion mass bilinear. At the chiral Heisenberg GNY QCP, the field content is $(N_\mathrm{f},N_\mathrm{b}^{\mathrm{QCP}})=(4,3)$, whereas at the chiral Ising GNY QCP it is $(N_\mathrm{f},N_\mathrm{b}^{\mathrm{QCP}})=(2,1)$~\cite{Herbut2006PRL,Li2015njp}. In the initial AFM state at $U_0$, the order parameter still has three real components, but only the two transverse Goldstone modes are gapless and contribute to the universal corner logarithm; thus $N_\mathrm{b}^{U_0}\equiv N_{\mathrm{b,gapless}}^{U_0}=N_G=2$, while the longitudinal Higgs mode and the fermions are gapped. In the initial CDW state at $V_0$, the Ising amplitude mode and the fermions are both gapped, so $N_\mathrm{b}^{V_0}\equiv N_{\mathrm{b,gapless}}^{V_0}=N_G=0$. The geometry-dependent free-field corner coefficients are obtained by summing the entries in Table~\ref{tab_sc_species} over all corner angles and multiplying by the appropriate number of gapless species.

\begin{table}[h]
	\centering
	\begin{tabular}{@{}l|ccc@{}}
		\hline\hline
		\addlinespace[0.2em]
		& $\theta=\frac{\pi}{3}$ & $\theta=\frac{2\pi}{3}$ & $\theta=\frac{\pi}{2}$ \\
		\addlinespace[0.2em]
		\hline
		\addlinespace[0.2em]
		$a_2^{\mathrm{f}}(\theta)$ & $0.0330$ & $0.0059$ & $0.01496$ \\
		\addlinespace[0.2em]
		\hline
		\addlinespace[0.2em]
		$a_2^{\mathrm{b}}(\theta)$ & $0.01465$ & $0.00255$ & $0.0065$ \\
		\hline\hline
	\end{tabular}
	\caption{Corner entanglement coefficients $a_2^{\mathrm{f}}(\theta)$ and $a_2^{\mathrm{b}}(\theta)$ for a single Dirac fermion and a single real boson at the second R\'enyi index~\cite{Helmes2016PRB}. The values of $a_2^{\mathrm{b}}$ refer to a single real boson, i.e., half of the corresponding complex-boson values. The total SCEE corner coefficient for a given model is obtained by summing over all corner angles of the subsystem geometry and multiplying by the number of species. For the honeycomb lattice, $s_c = N_{\mathrm{f/b}}\times 2[a_2(\frac{\pi}{3})+a_2(\frac{2\pi}{3})]$; for the $\pi$-flux square lattice, $s_c = N_{\mathrm{f/b}}\times 4\,a_2(\frac{\pi}{2})$.}
	\label{tab_sc_species}
\end{table}

\section{II. Projector quantum Monte Carlo and R\'enyi entropy}

We employ projector determinantal QMC (PQMC)~\cite{Sorella1989EPL,AssaadReview} to evaluate the R\'enyi entropy of a state generated by imaginary-time driving. For a generic interacting Hamiltonian $H[g(\tau)]=H_t+H_I[g(\tau)]$, PQMC starts from a trial Slater determinant and projects it in imaginary time. In the present calculation, this projection prepares the initial state at coupling $g_0$,
\begin{equation}
	\ket{\psi(0)} = \lim_{\tau_0\to\infty}e^{-\tau_0 H(g_0)}\ket{\Psi_T},
	\label{SM_PQMC_proj}
\end{equation}
The driven state is generated by the ordered imaginary-time evolution
\begin{equation}
	\ket{\psi(\tau_q)}=\mathcal{U}_R(\tau_q,0)\ket{\psi(0)},\qquad
	\mathcal{U}_R(\tau_q,0)=\mathrm{T}\exp\!\left[-\int_0^{\tau_q}d\tau\,H[g(\tau)]\right],
	\label{SM_driven_propagator}
\end{equation}
with $g(\tau)=g_0+\operatorname{sgn}(g_c-g_0)R\tau$ and $R=|g_c-g_0|/\tau_q$. Thus $g(\tau_q)=g_c$. The two sides of the PQMC projector contain a static preparation segment and a ramp segment with opposite time order. This is essential: Hamiltonians at different points on the ramp do not commute, so the driven state cannot be represented by an equilibrium projector at an averaged coupling~\cite{Zeng2025NC}.

We discretize both segments with $\Delta\tau=0.03$. Denoting $g_\ell=g(\ell\Delta\tau)$ and $N_q\Delta\tau=\tau_q$, the right branch of the contour is
\begin{equation}
	\mathbf{B}^{R}_{\mathbf{s}^R}=
	\underbrace{\prod_{\ell=N_q}^{1}e^{-\Delta\tau\mathbf{K}}
	e^{\mathbf{V}(\mathbf{s}_{\ell}^{R};g_\ell)}}_{\text{driven evolution}}
	\underbrace{\prod_{\ell=N_0}^{1}e^{-\Delta\tau\mathbf{K}}
	e^{\mathbf{V}(\mathbf{s}_{0,\ell}^{R};g_0)}}_{\text{initial-state projection}},
	\label{SM_Trotter}
\end{equation}
where the rightmost operator acts first. The left branch $\mathbf{B}^{L}_{\mathbf{s}^L}$ is the independently decoupled, anti-time-ordered Hermitian-conjugate contour. The initial projection length is chosen to converge the corresponding ground state, and the Trotter error at the above time step is negligible on the scale of the quoted statistical errors. Model-specific decouplings, projection lengths, and sign-free conditions are summarized in Sec.~III.

After a Hubbard--Stratonovich (HS) decoupling at each time slice, $\mathbf{V}(\mathbf{s}_{\ell};g_\ell)$ in Eq.~(\ref{SM_Trotter}) is bilinear in the fermions. The resulting auxiliary-field measure and the sign-free symmetry are model dependent, but the driven contour, endpoint Green's functions, and replica estimator below are common to all of the models studied here; their model-specific forms are given in Sec.~III.

The normalization of the driven state is sampled on the complete two-sided contour,
\begin{equation}
	\mathcal{Z}_{R}=\sum_{\mathbf{s}^{L},\mathbf{s}^{R}}
	\mathcal{W}_{\mathbf{s}^{L},\mathbf{s}^{R}},\qquad
	\mathcal{W}_{\mathbf{s}^{L},\mathbf{s}^{R}}=
	\mu(\mathbf{s}^{L},\mathbf{s}^{R})
	\det\!\left[\mathbf{P}^{\dagger}\mathbf{B}^{L}_{\mathbf{s}^{L}}
	\mathbf{B}^{R}_{\mathbf{s}^{R}}\mathbf{P}\right],
	\label{SM_partition}
\end{equation}
where $\mathbf{P}$ represents $\ket{\Psi_T}$ and $\mu$ is the product of the local HS measures on both branches. The equal-time Green's function is evaluated at the physical endpoint $\tau_q$, i.e., at the junction of the two branches,
\begin{equation}
	\mathbf{G}_{\mathbf{s}^{L},\mathbf{s}^{R}}(\tau_q)=\mathbb{I}-\mathbf{B}^{R}_{\mathbf{s}^{R}}\mathbf{P}
	\left[\mathbf{P}^{\dagger}\mathbf{B}^{L}_{\mathbf{s}^{L}}\mathbf{B}^{R}_{\mathbf{s}^{R}}\mathbf{P}\right]^{-1}
	\mathbf{P}^{\dagger}\mathbf{B}^{L}_{\mathbf{s}^{L}}.
	\label{SM_Green_eq}
\end{equation}
Although this junction is the midpoint of the full projector contour, it represents the final, driven state rather than an equilibrium midpoint. Time-displaced Green's functions connecting each update slice to $\tau_q$ are propagated with the actual ordered products in Eq.~(\ref{SM_Trotter}); no time-translation invariance or instantaneous-ground-state approximation is assumed.

For each drive rate, two independent replicas of the complete driven contour are sampled. At their common endpoint, the replica identity~\cite{Grover2013PRL}
\begin{equation}
	e^{-S_2^A}=\Tr[\rho_A^2]=\frac{Z_2}{Z_1^2}
	=\left\langle\det\mathbf{g}^A\right\rangle
	\label{SM_Renyi_identity}
\end{equation}
is evaluated from the Grover matrix restricted to the subsystem $A$,
\begin{equation}
	\mathbf{g}^A_{\mathbf{s}_1,\mathbf{s}_2} = \mathbf{G}^A_{\mathbf{s}_1}(\tau_q)\,\mathbf{G}^A_{\mathbf{s}_2}(\tau_q) + [\mathbb{I} - \mathbf{G}^A_{\mathbf{s}_1}(\tau_q)][\mathbb{I} - \mathbf{G}^A_{\mathbf{s}_2}(\tau_q)].
	\label{SM_Grover}
\end{equation}
Thus both the state preparation and the full noncommuting ramp enter the entanglement estimator through the endpoint Green's functions. This construction lets us measure the corner entropy directly at the target QCP for every $R$, instead of inferring it from a sequence of equilibrium states.

Local updates remain efficient despite the nonuniform time dependence. For a proposed change $s_x\to s_x'$ at the space--time point $x=(i,\ell)$, define the rank-1 update matrix in each independent fermion block $a$ as $\mathbf{r}_{x}^{(a)}=\mathbb{I}+\Delta_x^{(a)}[\mathbb{I}-\mathbf{G}_{\mathbf{s}}^{(a)}(\tau_q)]$. The full physical configuration-weight ratio is
\begin{equation}
	r = \frac{\mathcal W_{\mathbf s'}}{\mathcal W_{\mathbf s}}
	=\frac{w_x(s_x')}{w_x(s_x)}
	\prod_{a}\det\mathbf{r}_{x}^{(a)},
	\label{SM_acceptance}
\end{equation}
and the Metropolis acceptance probability is $\min(1,r)$ for a symmetric proposal. Here $w_x$ is the local HS measure and $\Delta_{x}^{(a)}$ denotes the rank-1 update matrix. The time-displaced propagators communicate a local change at any point on either the preparation or ramp segment to the endpoint without rebuilding the full contour. In the following equations, $\mathbf r$ denotes the matrix $\mathbf r_x^{(a)}$ for the updated block:
\begin{equation}
	\left\{
	\begin{aligned}
		\mathbf{G}_{\mathbf{s}_{1}^\prime}(\tau_q) &= \mathbf{G}_{\mathbf{s}_{1}}(\tau_q) - \mathbf{G}_{\mathbf{s}_{1}}(\tau_q)\mathbf{r}^{-1}\Delta[\mathbb{I}-\mathbf{G}_{\mathbf{s}_{1}}(\tau_q)] ,\\
		\mathbf{G}_{\mathbf{s}_1^\prime}(\tau) &= \mathbf{G}_{\mathbf{s}_1}(\tau) + \mathbf{G}_{\mathbf{s}_1}(\tau,\tau_q)\mathbf{r}^{-1}\Delta \mathbf{G}_{\mathbf{s}_1}(\tau_q,\tau), \\
		\mathbf{G}_{\mathbf{s}_1^\prime}(\tau_q,\tau) &= \mathbf{G}_{\mathbf{s}_1}(\tau_q,\tau) + \mathbf{G}_{\mathbf{s}_1}(\tau_q)\mathbf{r}^{-1}\Delta \mathbf{G}_{\mathbf{s}_1}(\tau_q,\tau), \\
		\mathbf{G}_{\mathbf{s}_{1}^\prime}(\tau,\tau_q) &= \mathbf{G}_{\mathbf{s}_{1}}(\tau,\tau_q) - \mathbf{G}_{\mathbf{s}_{1}}(\tau,\tau_q)\mathbf{r}^{-1}\Delta[\mathbb{I}-\mathbf{G}_{\mathbf{s}_{1}}(\tau_q)].
	\end{aligned}
	\right.
	\label{SM_Green_update}
\end{equation}
Similarly, the Grover matrix $\mathbf{g}^A_{\mathbf{s}_1, \mathbf{s}_2}$ can be updated efficiently. An update in the configuration of the first replica, $\mathbf{s}_1 \to \mathbf{s}_1^\prime$, results in the relation $\mathbf{g}^A_{\mathbf{s}_1^\prime, \mathbf{s}_2} (\mathbf{g}^A_{\mathbf{s}_1, \mathbf{s}_2})^{-1} = \mathbb{I} + \hat{a} \hat{b}$, with:
\begin{equation}
	\hat{a} = \mathbf{G}_{\mathbf{s}_1}(\tau, \tau_q) \mathbf{r}^{-1}\Delta, \quad
	\hat{b} = \mathbf{G}_{\mathbf{s}_1}(\tau_q, \tau) [2\mathbf{G}_{\mathbf{s}_2}(\tau) - \mathbb{I}] (\mathbf{g}^{A}_{\mathbf{s}_1, \mathbf{s}_2})^{-1}.
	\label{SM_Grover_update1}
\end{equation}
An analogous expression holds for updates to the second replica, $\mathbf{s}_2 \to \mathbf{s}_2^\prime$:
\begin{equation}
	\hat{a} = [2\mathbf{G}_{\mathbf{s}_1}(\tau) - \mathbb{I}] \mathbf{G}_{\mathbf{s}_2}(\tau, \tau_q), \quad
	\hat{b} = \mathbf{r}^{-1}\Delta \mathbf{G}_{\mathbf{s}_2}(\tau_q, \tau) (\mathbf{g}_{\mathbf{s}_1, \mathbf{s}_2}^{A})^{-1}.
	\label{SM_Grover_update2}
\end{equation}
By exploiting the rank-1 structure of $\Delta$, only a single row or column slice of $\hat{a}$ and $\hat{b}$ is required. The endpoint Grover determinant is therefore updated in $\mathcal{O}(N^2)$ operations rather than recomputed in $\mathcal{O}(N^3)$ operations after every field flip. Together with the incremental SCEE reweighting described in Sec.~IV, this makes it practical to resolve the small driven corner contribution while retaining the complete imaginary-time history of the ramp.

\section{III. Hubbard--Stratonovich transformations and sign-problem-free symmetries}

The contour and replica construction of Sec.~II requires only a bilinear HS vertex. Here we specify the model-dependent decouplings and the symmetries that make their configuration weights non-negative. For the half-filled $t$-$U$ Hubbard model, we use the four-valued, SU(2)-symmetric transformation
\begin{equation}
	e^{-\Delta\tau \frac{U(\tau_\ell)}{2}(n_{i,\uparrow}+n_{i,\downarrow}-1)^{2}} = \frac{1}{4} \sum_{s_{i} \in \{\pm 1, \pm 2\}} \gamma(s_{i})\, e^{i \alpha_\ell \eta(s_{i}) (n_{i,\uparrow}+n_{i,\downarrow}-1)} + \mathcal{O}(\Delta_{\tau}^{4}),
	\label{SM_HS}
\end{equation}
where $\alpha_\ell=\sqrt{\Delta_\tau U(\tau_\ell)/2}$, $\gamma(\pm 1)=1+\sqrt{6}/3$, $\gamma(\pm 2)=1-\sqrt{6}/3$, $\eta(\pm 1)=\pm\sqrt{2(3-\sqrt{6})}$, and $\eta(\pm 2)=\pm\sqrt{2(3+\sqrt{6})}$. Thus the local measure in Eq.~(\ref{SM_acceptance}) is $w_x(s_x)=\gamma(s_x)/4$. Under the partial particle--hole transformation $c_{i\downarrow}\to\epsilon_i c_{i\downarrow}^{\dagger}$, where $\epsilon_i=+1$ ($-1$) on sublattice $A$ ($B$), the HS-coupled density transforms as $n_{i\uparrow}+n_{i\downarrow}-1\to n_{i\uparrow}-n_{i\downarrow}$. In the real bipartite hopping gauge, both the hopping and HS-decoupled terms are invariant under the antiunitary transformation $\mathcal{T}=i\sigma_yK$, with $\mathcal{T}^2=-1$. The fermion eigenvalues therefore occur in complex-conjugate pairs, and the PQMC determinant is non-negative for every HS configuration~\cite{Wu2005PRBsign}.

For the spinless $t$-$V$ model, the density-channel decomposition in the complex-fermion representation is replaced by an HS decomposition in a Majorana hopping channel. We introduce two Majorana fermions at each site using the sublattice-dependent convention
\begin{equation}
	c_i=\begin{cases}
		\frac12(\gamma_i^2+i\gamma_i^1),&i\in A,\\[3pt]
		\frac12(\gamma_i^1+i\gamma_i^2),&i\in B.
	\end{cases}
	\label{SM_Majorana_transform}
\end{equation}
With each nearest-neighbor bond included once in the oriented set $\langle ij\rangle_{\rm or}$, the hopping term becomes
\begin{equation}
	H_t=\frac{it}{2}\sum_{\langle ij\rangle_{\rm or}}\zeta_{ij}
	\left(\gamma_i^1\gamma_j^1-\gamma_i^2\gamma_j^2\right),
	\qquad \zeta_{ij}=\pm1.
	\label{SM_tV_hopping}
\end{equation}
Here $\zeta_{ij}=1$ on the honeycomb lattice, whereas $\prod_\square\zeta_{ij}=-1$ encodes the $\pi$ flux on the square lattice. In the same convention, the interaction is
\begin{equation}
	H_V=\frac{V}{2}\sum_{\langle ij\rangle}
	\left(\frac{\gamma_i^1\gamma_j^1-\gamma_i^2\gamma_j^2}{2}\right)^2
	+\frac{V N_{\rm bond}}{4},
	\label{SM_HV_Majorana}
\end{equation}
where $N_{\rm bond}$ is the number of nearest-neighbor bonds. With the four-valued coefficients in Eq.~(\ref{SM_HS}), the bond interaction is decoupled as
\begin{equation}
	e^{-\Delta\tau H_V}
	=e^{-\Delta\tau V N_{\rm bond}/4}
	\prod_{\langle ij\rangle}
	\left[\frac14\sum_{s_{ij}\in\{\pm1,\pm2\}}\gamma(s_{ij})\,
	\exp\!\left(i\sqrt{\frac{\Delta\tau V}{8}}\,\eta(s_{ij})
	\left(\gamma_i^1\gamma_j^1-\gamma_i^2\gamma_j^2\right)\right)\right]
	+\mathcal{O}(\Delta\tau^4).
	\label{SM_HS_Majorana}
\end{equation}
Equations~(\ref{SM_tV_hopping}) and (\ref{SM_HS_Majorana}) have the same Majorana structure. For every auxiliary-field configuration, the two Majorana sectors and their projected weights obey
\begin{equation}
	\begin{gathered}
		h^2(\mathbf{s}_\ell)=\left[h^1(\mathbf{s}_\ell)\right]^*,\qquad
		W_2(\mathbf{s})=W_1(\mathbf{s})^*,\\
		\mathcal W_{\mathbf{s}}^{tV}=
		\left[\prod_{\ell=1}^{N_\tau}\prod_{\langle ij\rangle}
		\frac{\gamma(s_{ij,\ell})}{4}\right]
		W_1(\mathbf{s})W_2(\mathbf{s})
		=\left[\prod_{\ell=1}^{N_\tau}\prod_{\langle ij\rangle}
		\frac{\gamma(s_{ij,\ell})}{4}\right]
		\left|W_1(\mathbf{s})\right|^2\ge0.
	\end{gathered}
\end{equation}
The full bilinear therefore belongs to the Majorana symmetry class $\{\mathcal{T}_1^+,\mathcal{T}_2^-\}$, where $(\mathcal{T}_i^\pm)^2=\pm1$. So the bond-field measure does not alter the sign-problem-free conclusion for the half-filled spinless $t$-$V$ model~\cite{Li2016PRLsign,li2019arcmp}. For a single bond-field update $s_{ij,\ell}\to s_{ij,\ell}'$, the full physical weight ratio analogously contains $\gamma(s_{ij,\ell}')/\gamma(s_{ij,\ell})$ in addition to the fermionic ratio.

\section{IV. Incremental algorithm for the SCEE}

The R\'enyi entropy $S_2$ exhibits extensive scaling with subsystem size, so the estimator $e^{-S_2}$ decays exponentially with system size $L$, causing a severe signal-to-noise ratio problem~\cite{Grover2013PRL}. To mitigate these exponentially growing relative statistical errors, we employ the incremental algorithm~\cite{Emidio2024PRL,Liao2025npjQI,DaliaoYuan2023PRB}. Rather than computing $S_2^A$ and $S_2^B$ separately, the ratio $e^{-\Delta S_2} = e^{-(S_2^A - S_2^B)}$ is sampled directly:
\begin{equation}
	e^{-\Delta S_2} = \frac{\sum_{\mathbf{s}_1,\mathbf{s}_2} P_{\mathbf{s}_1,\mathbf{s}_2}\det\mathbf{g}^A_{\mathbf{s}_1,\mathbf{s}_2}}{\sum_{\mathbf{s}_1,\mathbf{s}_2} P_{\mathbf{s}_1,\mathbf{s}_2}\det\mathbf{g}^B_{\mathbf{s}_1,\mathbf{s}_2}}.
	\label{SM_SCEE_ratio}
\end{equation}
Here $P_{\mathbf{s}_1,\mathbf{s}_2}\propto\mathcal W_{\mathbf{s}_1}\mathcal W_{\mathbf{s}_2}$ denotes the product of the full physical configuration weights of the two replicas, including the model-specific local HS measures specified in Sec.~III. For the spinless $t$-$V$ model, these weights are the nonnegative physical two-Majorana-sector weights defined in Sec.~III, and $\mathbf g$ is constructed from the physical complex-fermion Green's function. Thus, all fractional powers below act only on physical, nonnegative weight determinants.
The expectation value $O \equiv \det \mathbf{g}^A$ is decomposed into a product of incremental ratios along a monotonically increasing path $\boldsymbol{\lambda} = \{\lambda_0=0, \lambda_1, \ldots, \lambda_N=1\}$:
\begin{equation}
	\langle O \rangle = \prod_{n=0}^{N-1} \frac{\sum_i P_i O_i^{\lambda_{n}} O_i^{\Delta_{\lambda}}}{\sum_i P_i O_i^{\lambda_{n}}}.
	\label{SM_incremental}
\end{equation}

We adopt a linear parameterization $\lambda_n = n/N_\lambda$, where $\Delta_{\lambda} = 1/N_\lambda$ is the uniform incremental step size. As the lattice size $L$ increases, the number of intermediate $\lambda$-points must be increased to constrain the statistical error. The logarithm of the determinant, $\ln(\det \mathbf{g}^A)$, follows a normal distribution $\mathcal{N}(\mu, \sigma^2)$, with the mean and standard deviation scaling as $\mu(L) \propto L^\alpha$ and $\sigma(L) \propto L^\beta$. The coefficient of variation for the raw determinant, $\mathrm{CV}[\det \mathbf{g}^A] = \sqrt{e^{\sigma^2}-1}$, grows exponentially with $L$. In contrast, the coefficient of variation for the incremental step scales as $\mathrm{CV}[(\det \mathbf{g}^A)^{1/N_\lambda}] = \sqrt{e^{\sigma^2/N_\lambda^2}-1}$, which remains bounded provided $N_\lambda > \sigma(L)$. In practice, we select $N_\lambda(L) \approx 0.2 L^{1.7} > \sigma(L)$.

At each incremental step $n$, the effective sampling weight and observable are defined as:
\begin{equation}
	\left\{
	\begin{aligned}
		&\mathbf{P}_{i}^{(n)} \equiv P_i O_i^{\lambda_{n}}, \\
		&O_{i}^{(n)} \equiv O_i^{\Delta_{\lambda}}.
	\end{aligned}
	\right.
	\label{SM_incremental_weight}
\end{equation}
The modified acceptance ratio for a single auxiliary-field flip at step $n$ is $R_{\lambda_n} = r \cdot (\det \Gamma_A)^{\lambda_n}$, where $r$ is the full physical configuration-weight ratio of Eq.~(\ref{SM_acceptance}) and already contains the appropriate local HS-measure factor $w_x(s_x')/w_x(s_x)$. The matrix $\Gamma_A \equiv \mathbb{I} + \hat{b}\hat{a}$ accounts for the subsystem contribution from the Grover matrix update.

In the combined incremental and SCEE (ICR-SCEE) method~\cite{kups7666,Xu2024PRLDisorderoperator,DaLiaoYuan2024PRB}, the interpolation is performed between the two subsystem geometries $A$ and $B$ rather than between the trivial state and a single subsystem. The effective weight and observable for the $n$-th incremental step are:
\begin{equation}
	\left\{
	\begin{aligned}
		\mathcal{P}_{i}^{(n)} &\equiv P_{\mathbf{s}_1,\mathbf{s}_2} (\det \mathbf{g}^B_{\mathbf{s}_1,\mathbf{s}_2})^{1-\lambda_n} (\det \mathbf{g}^A_{\mathbf{s}_1,\mathbf{s}_2})^{\lambda_n}, \\
		\mathcal{O}_{i}^{(n)} &\equiv \left( \frac{\det \mathbf{g}^A_{\mathbf{s}_1,\mathbf{s}_2}}{\det \mathbf{g}^B_{\mathbf{s}_1,\mathbf{s}_2}} \right)^{\Delta_{\lambda}}.
	\end{aligned}
	\right.
	\label{SM_ICRSCEE_weight}
\end{equation}
This formulation ensures a smooth interpolation between the two subsystem geometries. The corresponding ICR-SCEE update ratio for an auxiliary-field flip is:
\begin{equation}
	R_{\lambda_n} = r \cdot \det(\Gamma_B)^{1-\lambda_n} \cdot \det(\Gamma_A)^{\lambda_n},
	\label{SM_ICRSCEE_ratio}
\end{equation}
where $\Gamma_{A/B}$ represent the subsystem-dependent update factors for $\det \mathbf{g}^{A/B}$ derived from the rank-1 matrix modifications in Eqs.~(\ref{SM_Grover_update1})--(\ref{SM_Grover_update2}). The same full ratio $r$ defined above is used here, so the local HS-measure factor is already included and must not be multiplied a second time. This factorization into many small ratios dramatically reduces the variance compared to direct sampling of the full ratio, enabling high-precision measurement of $\Delta S_2$.

\section{V. Extraction of the corner coefficient at the QCP}

A useful consequence of Eq.~(\ref{scaling_main}) is that it enables the
universal QCP corner coefficient $s_c^{\rm QCP}$ to be extracted from
the driving-rate dependence of $\Delta S_2$. Within the low-energy FTS
window $RL^r\gg1$, but below the microscopic fast-driving regime,
Eq.~(\ref{scaling_main}) predicts a linear dependence on $\ln R$ with
$\mathrm{slope}=\frac{s_c^0-s_c^{\rm QCP}}{r}$, 
where $s_c^0$ is the corner coefficient of the initial ordered state and
$r=z+1/\nu$. Using the correlation-length exponent $\nu$ obtained from
independent studies of local order parameters
~\cite{Li2015njp,Sorella2016prx,wang2026}, together with $z=1$, the QCP
corner coefficient is extracted as
\begin{equation}
    s_c^{\rm QCP}=s_c^0-r\,\mathrm{slope}.
    \label{SM_sc_result}
\end{equation}

The corresponding uncertainty is obtained by standard error propagation:
\begin{equation}
    \delta s_c^{\rm QCP}
    =
    \sqrt{
        (\delta s_c^0)^2
        +r^2[\delta(\mathrm{slope})]^2
        +(\mathrm{slope})^2\frac{(\delta\nu)^2}{\nu^4}
    }.
    \label{SM_error_prop}
\end{equation}
The three terms account for the uncertainty in the initial-state
coefficient, the statistical uncertainty of the fitted slope, and the
uncertainty in $\nu$, respectively. When $s_c^0$ is used as an exact
theoretical input, the first term is set to zero.

The extracted values of $s_c^{\rm QCP}$ for all four models are listed
in Table~\ref{tab_models}. For comparison, the table also includes the
free-fermion contribution. These free-field values provide useful reference scales. And the dynamically extracted results are also consistent with previous entanglement-based estimates~\cite{shen2026}, supporting the validity of the present extraction procedure.

\begin{figure}[t]
	\centering
		\includegraphics[width=0.8\linewidth]{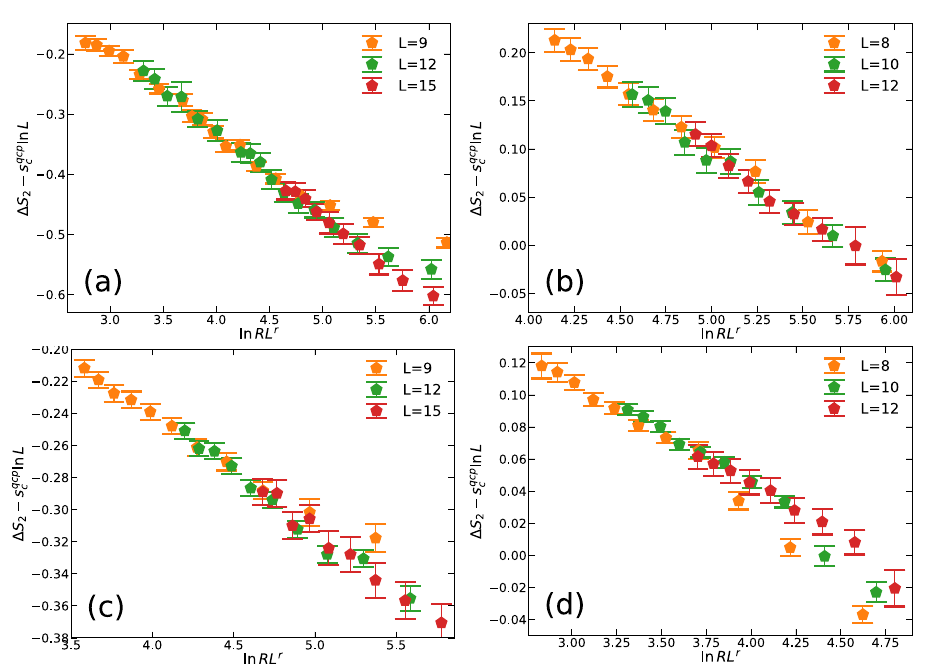}
	\caption{Data collapse of the driven scaling function $f(x)$ of Eq.~(\ref{SM_collapse_form}). The subtracted corner entanglement entropy $\Delta S_2-s_c^{\rm QCP}\ln L$ is plotted against the scaling variable $\ln(RL^r)$, with $r=z+1/\nu$. (a) Hubbard model on the honeycomb lattice. (b) Hubbard model on the $\pi$-flux square lattice. (c) Spinless $t$-$V$ model on the honeycomb lattice. (d) Spinless $t$-$V$ model on the $\pi$-flux square lattice. For finite driving rates, data for different $L$ and $R$ collapse onto a single straight line of slope $(s_c^0-s_c^{\rm QCP})/r$, whereas data at the extreme rates $R\to0$ and $R\to\infty$ depart from the universal curve, where the scaling theory no longer applies.}
	\label{SM_collapse}
\end{figure}

\section{VI. Data collapse of the driven scaling function \texorpdfstring{$f(x)$}{f(x)}}

To verify the driven scaling form directly, we test the scaling function $f(x)$ in
\begin{equation}
	\Delta S_2(R,L) = s_c^{\rm QCP}\ln L + f(RL^r) + \text{const.},
	\label{SM_collapse_form}
\end{equation}
where $r=z+1/\nu$. The full scaling function $f(RL^r)$ interpolates
between equilibrium finite-size scaling and the FTS regime. For
$RL^r\ll1$, corresponding to $\xi_R\gg L$, $f(RL^r)$ approaches a
constant, and Eq.~(\ref{SM_collapse_form}) recovers the equilibrium
result. For $RL^r\gg1$, provided that the driving rate remains within
the low-energy scaling window $a\ll\xi_R\ll L$, the asymptotic form is
\[
f(RL^r)=\frac{s_c^0-s_c^{\rm QCP}}{r}\ln(RL^r)+\text{const.}
\]
Consequently, plotting $\Delta S_2-s_c^{\rm QCP}\ln L$ against
$\ln(RL^r)$ yields a straight line as shown in Fig.~\ref{SM_collapse}.
Deviations at small $R$ reflect the crossover from this logarithmic
asymptote to equilibrium finite-size scaling, rather than a breakdown
of the full scaling function. At excessively large $R$, microscopic
high-energy excitations can instead generate nonuniversal corrections
to the low-energy scaling behavior~\cite{Zeng2025NC}.

\end{document}